\documentclass[twocolumn]{aastex61}
\pdfoutput=1
\usepackage{silence}
\usepackage{graphicx}
\graphicspath{{images/}}
\usepackage[caption=false]{subfig}
\usepackage{booktabs}
\usepackage{stmaryrd}

\usepackage{natbib}
\setcitestyle{authoryear,open={(},close={)}}

\usepackage{float}
\usepackage{color,soul}
\usepackage{wrapfig}
\usepackage{bm}
\usepackage{amsmath}

\usepackage{fancyhdr}
\usepackage{wasysym}

\newcommand{\lir}{L$_{\mathrm{IR}}$}

\newcommand{\loglir}{$\log\mathrm{L}_{\mathrm{IR}}/\mathrm{L}_\odot$}

\begin{document}

\title{Broad emission lines in optical spectra of hot dust-obscured galaxies can contribute significantly to JWST/NIRCam photometry}
\author[0000-0002-6149-8178]{Jed McKinney}
\affiliation{Department of Astronomy, The University of Texas at Austin, 2515
Speedway Blvd Stop C1400, Austin, TX 78712, USA}

\author[0000-0002-1392-0768]{Luke Finnerty}
\affiliation{Department of Physics \& Astronomy, 430 Portola Plaza, University of California, Los Angeles, CA 90095, USA}

\author[0000-0002-0930-6466]{Caitlin M. Casey}
\affiliation{Department of Astronomy, The University of Texas at Austin, 2515
Speedway Blvd Stop C1400, Austin, TX 78712, USA}

\author[0000-0002-3560-8599]{Maximilien Franco}
\affiliation{Department of Astronomy, The University of Texas at Austin, 2515
Speedway Blvd Stop C1400, Austin, TX 78712, USA}

\author[0000-0002-7530-8857]{Arianna S. Long}
\altaffiliation{NASA Hubble Fellow}
\affiliation{Department of Astronomy, The University of Texas at Austin, 2515
Speedway Blvd Stop C1400, Austin, TX 78712, USA}

\author[0000-0001-7201-5066]{Seiji Fujimoto}
\altaffiliation{NASA Hubble Fellow}
\affiliation{Department of Astronomy, The University of Texas at Austin, 2515
Speedway Blvd Stop C1400, Austin, TX 78712, USA}
\affiliation{Cosmic Dawn Center (DAWN), Jagtvej 128, DK2200 Copenhagen N, Denmark}
\affiliation{Niels Bohr Institute, University of Copenhagen, Lyngbyvej 2, DK2100 Copenhagen \O, Denmark}

\author[0000-0002-7051-1100]{Jorge A. Zavala}
\affiliation{National Astronomical Observatory of Japan, 2-21-1 Osawa, Mitaka, Tokyo 181-8588, Japan}

\author[0000-0003-3881-1397]{Olivia Cooper}
\affiliation{Department of Astronomy, The University of Texas at Austin, 2515
Speedway Blvd Stop C1400, Austin, TX 78712, USA}

\author[0000-0003-3596-8794]{Hollis Akins}
\affiliation{Department of Astronomy, The University of Texas at Austin, 2515
Speedway Blvd Stop C1400, Austin, TX 78712, USA}

\author[0000-0001-8592-2706]{Alexandra Pope} 
\affiliation{Department of Astronomy, University of Massachusetts, Amherst, MA 01003, USA.}

\author[0000-0003-3498-2973]{Lee Armus}
\affiliation{IPAC, California Institute of Technology, 1200 E. California Blvd., Pasadena, CA 91125, USA}

\author[0000-0002-8112-1132]{B. T. Soifer}
\affiliation{Division of Physics, Math, and Astronomy, California Institute of Technology, 1200 E California Blvd., Pasadena, CA 91125, USA}

\author[0000-0003-3917-6460]{Kirsten Larson}
\affiliation{AURA for the European Space Agency (ESA), Space Telescope Science Institute, 3700 San Martin Drive, Baltimore, MD 21218, USA}

\author{Keith Matthews}
\affiliation{Division of Physics, Math, and Astronomy, California Institute of Technology, 1200 E California Blvd., Pasadena, CA 91125, USA}

\author{Jason Melbourne}
\affiliation{Division of Physics, Math, and Astronomy, California Institute of Technology, 1200 E California Blvd., Pasadena, CA 91125, USA}

\author[0000-0001-7780-3352]{Michael Cushing}
\affiliation{Department of Physics and Astronomy, University of Toledo, 2801 West Bancroft St., Toledo, OH 43606, USA}

%\author{collaborators}

\begin{abstract}
Selecting the first galaxies at $z>7-10$ from \textit{JWST} surveys is complicated by $z<6$ contaminants with degenerate photometry. For example, strong optical nebular emission lines at $z<6$ may mimic \textit{JWST}/NIRCam photometry of $z>7-10$ Lyman Break Galaxies (LBGs). Dust-obscured $3<z<6$ galaxies in particular are potentially important contaminants, and their faint rest-optical spectra have been historically difficult to observe. 
A lack of optical emission line and continuum measures for $3<z<6$ dusty galaxies now makes it difficult to test their expected \textit{JWST}/NIRCam photometry for degenerate solutions with NIRCam dropouts.
Towards this end, we quantify the contribution by strong emission lines to NIRCam photometry in a physically motivated manner by stacking 21 Keck II/NIRES spectra of hot, dust-obscured, massive ($\log\mathrm{M_*/M_\odot}\gtrsim10-11$) and infrared (IR) luminous galaxies at $z\sim1-4$. We derive an average spectrum and measure strong narrow (broad) [OIII]$_{5007}$ and H$\alpha$ features with equivalent widths of $130\pm20\,$\AA\ ($150\pm50\,$\AA) and $220\pm30\,$\AA\ ($540\pm80\,$\AA) respectively. 
These features can increase broadband NIRCam fluxes by factors of $1.2-1.7$ ($0.2-0.6\,$mag). 
Due to significant dust-attenuation ($A_V\sim6$), we find H$\alpha$+[NII] to be significantly brighter than [OIII]+H$\beta$, and therefore find that emission-line dominated contaminants of high$-z$ galaxy searches can only reproduce moderately blue perceived UV continua of $S_\lambda\propto\lambda^\beta$ with $\beta>-1.5$ and $z>4$. 
While there are some redshifts ($z\sim3.75$) where our stack is more degenerate with the photometry of $z>10$ LBGs between $\lambda_{rest}\sim0.3-0.8\,\mu$m, redder filter coverage beyond $\lambda_{obs}>3.5\,\mu$m and far-IR/sub-mm follow-up may be useful for breaking the degeneracy and making a crucial separation between two fairly unconstrained populations, dust-obscured galaxies at z$\sim3-6$ and LBGs at z$>10$. 
\end{abstract}

%\begin{itemize}
%\item \sout{Fill in missed PAH 6.2 um luminosity for GS IRS70 and GS IRS81.} 
%\item \sout{re-visit dust mass calculation. Am I actually using the WG2001 opacity tables?}
%\item \sout{ Run imagetool code to get map RMS for each target and put those numbers into Table 2. }
%\item \sout{Correct the axis labels specifically for \sigmaIR\ which seem in some cases to be missing some text. Also, add the limits to the GOALS galaxies for IR sizes on these figures.}
%\item \sout{Add a table that has all of the linear relation parameters }
%\item \sout{Add in the postage stamp figure}
%\item fill in conclusion text 
%\item Fill out the abstract
%\end{itemize}

\section{Introduction}
%\begin{itemize}
    %\item Dusty galaxies harbor most of the star-formation in the Universe at the star-formation rate density peak. 
    %\item Beyond $z\sim1-3$, it is unclear precisely how prevalent this population is. Nevertheless, number counts suggest a massive population is in place at $z>3$.
    %\item JWST opens up the possibility of detecting the rest-frame optical emission of such dusty galaxies. This opens the door for their selection in upcoming deep extragalctic surveys (CEERS, COSMOS-Web, PRIMER). 
    %\item Nebular line contribution to broadband photometry: %\cite{Zackrisson2008,Schaerer2009,Stark2013,Wilkins2013,Wilkins2020,Wilkins2022}.. 
%\end{itemize}

A major objective baked into the design of \textit{JWST} is detecting the light from the first galaxies residing at ultra-high redshifts ($z>10$). Delivering on its promise, more than 30 galaxy candidates with photometric redshift solutions favoring $z>10$ were identified within the first months of publicly available data \citep{Bradley2022,Donnan2022,Naidu2022,Harikane2022,Finkelstein2022,Hsiao2022}. Assessing the fidelity of these samples is critical, particularly because the statistics assuming current $z>10-15$ candidates are real may or may not violate $\Lambda$CDM predictions  \citep{Naidu2022,BoylanKolchin2022,Labbe2022,Maio2022}. 

Spectroscopic confirmation is needed to verify these redshifts. However, some early attempts at spectroscopic follow-up using facilities such as ALMA have yielded upper limits or tentative low-SNR detections \citep[e.g.,][]{Bakx2022,Kaasinen2022,Yoon2022,Fujimoto2022}. \textit{JWST}/NIRSpec has proven capable of spectroscopically detecting the rest-frame optical light from galaxies up to $z\sim9-10$ \citep{Carnall2022,Roberts2022}, but this might not be well-suited for rapidly validating redshifts in statistical samples. A complimentary approach, born from similar rest-frame optical colors of $z>10$ Lyman Break Galaxies (LBGs) and dusty galaxies \citep{Howell2010,Casey2014b}, is to use far-IR/sub-mm followup observations of cold dust continuum \citep{Zavala2022} and/or far-IR cooling lines \citep{Fujimoto2022} to identify or rule out $z<6$ IR-luminous galaxies lurking within $z>10$ candidate catalogs. Dusty sources have posed a problem to the fidelity of high$-z$ galaxy catalogs since selection from \textit{Hubble Space Telescope} (\textit{HST}) extragalactic deep fields \citep[e.g.,][]{Dunlop2007}. \textit{HST} samples at $z\sim6-7$ were contaminated by $z\sim2$ dusty, star-forming galaxies. Now, $z\sim3-6$ dusty galaxies may be contaminating $z>7-10$ \textit{JWST} samples. A contributing factor to this contamination is that both populations have similar, and uncertain, number densities: dusty star-forming galaxies at $z\sim3-6$ have volume number densities $n\sim10^{-5}-10^{-6}\,\mathrm{Mpc^{-3}}$ \citep{Michalowski2017,Koprowski2017,RowanRobinson2018,Dudzeviciute2020,Gruppioni2020,Manning2022,Long2022}, similar to early measurements of bright $z>10$ LBGs \citep{Finkelstein2022,Naidu2022,Harikane2022,Bouwens2022b}. Disentangling these two populations is therefore also crucial for reducing uncertainties in their respective number densities, which are currently inflated by sample purity \citep[e.g.,][]{Bouwens2022b}.

%is capable of detecting the rest-frame optical light of these sources out to higher redshifts than \textit{HST}, which increases their likelihood of contaminating ultra-high redshift galaxy catalogs. 

%High signal-to-noise rest-frame optical/near-IR spectroscopy of IR-luminous galaxies are needed to identify the photometric signatures not present in $z>10$ LBGs. 
Of particular concern within the rest-frame optical/near-IR spectra of IR-luminous galaxies is the relative contribution of strong narrow and broad emission lines to broadband filter fluxes, which could mask red continuum slopes produced by dust attenuation. Strong nebular lines can change \textit{JWST}/NIRCam colors
\citep{Zackrisson2008,Schaerer2009,Stark2013,Wilkins2013,Wilkins2020,Wilkins2022}. This may be a promising tool for pseudo-spectroscopy of lower redshift galaxies using narrow-band filters, but in this particular instance is a source of potential population confusion for broadband high-redshift surveys. Indeed, some of the hottest and most luminous dusty galaxies at $z>1$ exhibit very high rest-frame optical line equivalent widths (EWs) \citep{Finnerty2020}. These arise from a combination of low dust-attenuated continuum levels with bright lines emergent from less obscured regions, as well as ionized outflows driven by Active Galactic Nuclei (AGN). To what extent do these strong lines contribute to \textit{JWST} photometry? 

In this letter, we take an empirically-grounded approach and quantify the contamination from emission lines from hot dust-obscured galaxies to \textit{JWST}/NIRCam photometry. % response to the rest-frame optical spectrum of dusty galaxies between $z=1-9$.
In Section \ref{sec:data} we describe Keck II/NIRES observations of a sample of 4 luminous, IR galaxies (\loglir$\sim12.5$) and 17 hot dust, obscured galaxies (DOGs, \loglir$\,>13$) between $z\sim1-4$, which we stack to derive an average optical spectrum (Section \ref{sec:analysis}). We quantify the contribution of strong and broad optical emission lines to NIRCam fluxes in Section \ref{sec:discussion}, and discuss their impact on distinguishing between such sources at $z<6$ and $z>10$ LBGs. Section \ref{sec:conclusion} summarizes our main findings. Throughout this work we adopt a $\Lambda$CDM cosmology with $\Omega_m=0.3,\,\Omega_\Lambda=0.7$ and $H_0=70\,\mathrm{km\,s^{-1}\,Mpc^{-1}}$.

%of interlopers to be dusty galaxies \citep{Dunlop2007}

%prove difficult for ruling out dusty galaxy solutions wi

%When fitting rest-frame optical/near-infrared photometry of Lyman Break Galaxy (LBG) candidates at $z\gtrsim6-10$, a lower redshift solution fit by the SED of a dusty galaxy can be allowed with near-equal likelihood \citep{Dunlop2007,Zavala2022}. In fact, far-infrared/sub-mm follow-up can be a useful tool in identifying the population of dusty interlopers within $z>10$ samples \citep[e.g.,][]{Zavala2022}. In the absence of far-infrared data, dusty galaxy solutions are often ruled out by their dust-attenuated red continuum slopes; however, strong rest-frame optical emission lines could in principle increase broad-band filter photometry, potentially masking the red continuum signal. 

\begin{figure*}[!ht]
    \centering
    \includegraphics[width=\textwidth]{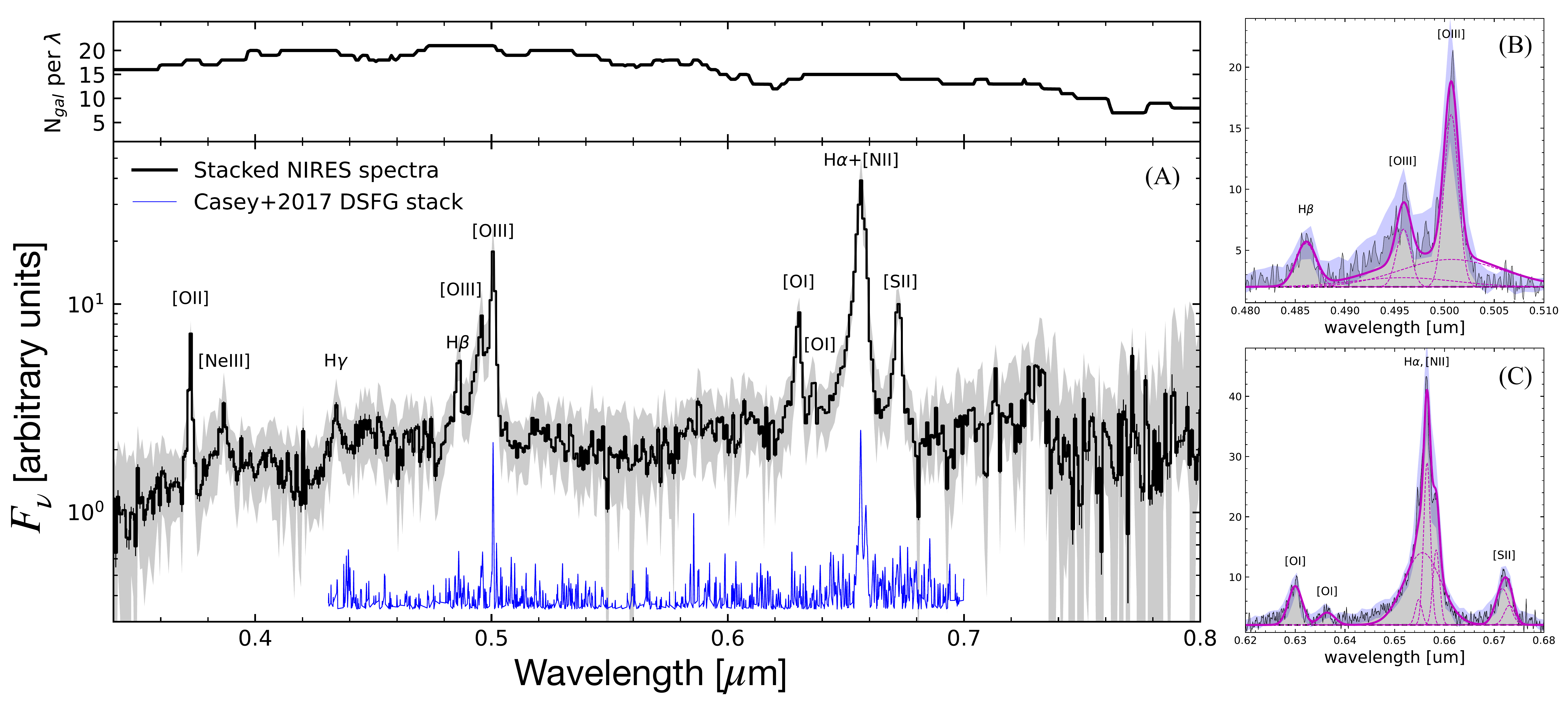}
    \caption{Stacked rest-frame optical spectrum of $z\sim1-4$ IR-luminous galaxies detected with Keck II/NIRES. (A) The mean-weighted stacked spectrum (black). 
    %, re-binned to a coarse resolution to improve the continuum SNR. 
    Shaded grey errors correspond to 16th-84th percentiles on the flux density derived from the boostrapped stack distribution per wavelength. 
    %Fits to the broad [O III] and H$\alpha$ features are shown (red dashed line) to distinguish between continuum and line emission. 
    The upper panel gives the number of galaxies included in the stack as a function of wavelength. On average, 70\% ($>14$) of the sample is represented between $\lambda_{rest}=0.34-0.80\,\mu$m. We compare against the stacked (N$=20$) continuum-normalized DSFG spectrum from \cite{Casey2017b} (blue). Panels (B) and (C) show zoom-ins on the [O III], H$\beta$ and [N II], H$\alpha$, [SII], [OI] features respectively. 
    %at the spectrum's native resolution, 
    We measure broad and narrow components as expected from the individual spectra \citep{Finnerty2020}. Line fits are shown in purple with solid lines indicating the total line$+$continuum fit and dashed lines for individual line profiles. 16th and 84th percentiles derived from 1000 bootstrapped stacks are shown in blue. %The shaded region encases the wavelength range over which we calculate EWs and mask from the line-free spectrum.
    }
    \label{fig:stack}
\end{figure*}

\section{Sample and Data\label{sec:data}}

Hot DOGs were originally selected from WISE photometry as W1W2-dropouts and include the most-luminous galaxies in the Universe \citep{Eisenhardt2012,Wu2012}. This extreme population is experiencing a rapid phase of both supermassive black-hole and stellar mass assembly \citep{Eisenhardt2012}, and are mostly found at $z\sim2-3$ with \loglir$\,\geq13$ \citep{Wu2012,Assef2015,Tsai2015}. % On-going star-formation and supermassive black-hole growth \citep{Eisenhardt2012}. 
Most hot DOGs exhibit strong ionized outflows in optical spectroscopy \citep{Wu2018,Finnerty2020,Jun2020}, as implied by broad line components likely driven by radiative AGN feedback \cite{Wu2018}.  
These sources are rare with only $\sim1000$ over the full WISE All-sky survey \citep{Cutri2012}.  %Rarer than typical sub-millimeter galaxies (SMGs) and dusty star-forming galaxies (DSFGs). Ionized outflows \citep{Wu2018,Finnerty2020,Jun2020}.

The Hot DOG spectra were previously described in \citet{Finnerty2020}. In brief, we obtained simultaneous $JHK$ spectra at $R\sim2700$ with Keck II/NIRES \citep{Wilson2004} and reduced the data using SPEXTOOL \citep{Cushing2004}. Flux calibration was performed by comparing the integrated flux with $K^\prime$ photometry, see \citet{Finnerty2020} for details. Our stacked spectrum uses the 17 sources with detections of [OIII], H$\beta$ and/or [NII], H$\alpha$. %We estimate the continuum spectrum by subtracting the best-fit [OIII], H$\beta$ and [NII], H$\alpha$, [SII] fits from \citet{Finnerty2020}. %No other lines were subtracted, though [SII] and [OII] features are well-detected in several hot DOGs. 

In addition to the hot DOGs, we include in our analysis previously unpublished Keck II/NIRES spectra of four $z\sim1-2$ galaxies with $\log\,\mathrm{M_*/M_\odot}\sim11$ and \loglir$\,\sim12.5$: 
GS 3 ($z=0.544$, RA/DEC = 03:32:08.66, -27:47:34.4), GS 7 ($z=1.042$, RA/DEC = 03:32:26.49, -27:40:35.7), GN 1 ($z=1.432$, RA/DEC = 12:36:45.83, +62:07:54.0), and GN 40 ($z=1.609$, RA/DEC = 12:36:49.65, +62:07:38.6). 
These targets were selected for existing \textit{Spitzer}/IRS mid-infrared spectroscopy and bright IRAC Ch.~1 photometry from a 24$\,\mu$m-selected parent sample \citep{Kirkpatrick2012,Kirkpatrick2015}. %, and are more typical with respect to SMGs. 
GS 3, GS 7, and GN 40 are mid-IR AGN \citep{Kirkpatrick2015}, and GN 1 is a composite source with a mid-IR AGN fraction of 50\%. These sources have \lir\ on-average an order of magnitude lower than those of the hot DOGs. We reduce the data for this sub-sample following the exact same procedures as the hot DOGs described in \citealt{Finnerty2020}. [O II], [OIII], H$\beta$ and [NII], H$\alpha$ are individually detected with EWs on-average lower than those of the hot DOGs but within their range.   

%The spectra of GS 7, GS 3, GN 40, and GN 1, which were not presented in \citet{Finnerty2020}, were obtained and analyzed following the same procedures as the Hot DOGs during a KECK/NIRES observing run in February, 2020. 

\section{Analysis\label{sec:analysis}}
In this work we compute synthetic photometry in broadband \textit{JWST}/NIRCam filters for the rest-frame optical spectrum of our stacked spectrum redshifted between $z=1-9$. %Individually, the continuum SNR of NIRES spectra are low so we stack hot DOGs and $z\sim1-2$ LIRGs to obtain an average rest-frame optical spectra for massive \loglir$\,>12$ galaxies at $z\sim1-4$. 
While we do not claim our sample to be definitively representative of \textit{all} dusty systems owing to the extreme nature of hot DOGs, the final stack is derived from empirical data with no modelling required. While a more detailed study exploring a range of continuum templates with added nebular lines is warranted, such analysis is beyond the scope of this work given the current lack of constraint on rest-frame optical spectra of dusty galaxies beyond $z>4-5$. To supplement our analysis of very luminous, massive hot DOGs, we also compute synthetic photometry for the average dusty, star-forming galaxy (DSFG) spectrum from \citealt{Casey2017b}. The \citealt{Casey2017b} stack is constructed from Keck/MOSFIRE spectra of 20 LIRGs and ULIRGs with $\langle z \rangle = 2.1$, a more typical IR-bright galaxy population selected from ground-based single dish sub-mm surveys \citep{Casey2013}.

\begin{figure*}[!ht]
    \centering
    \includegraphics[width=\textwidth]{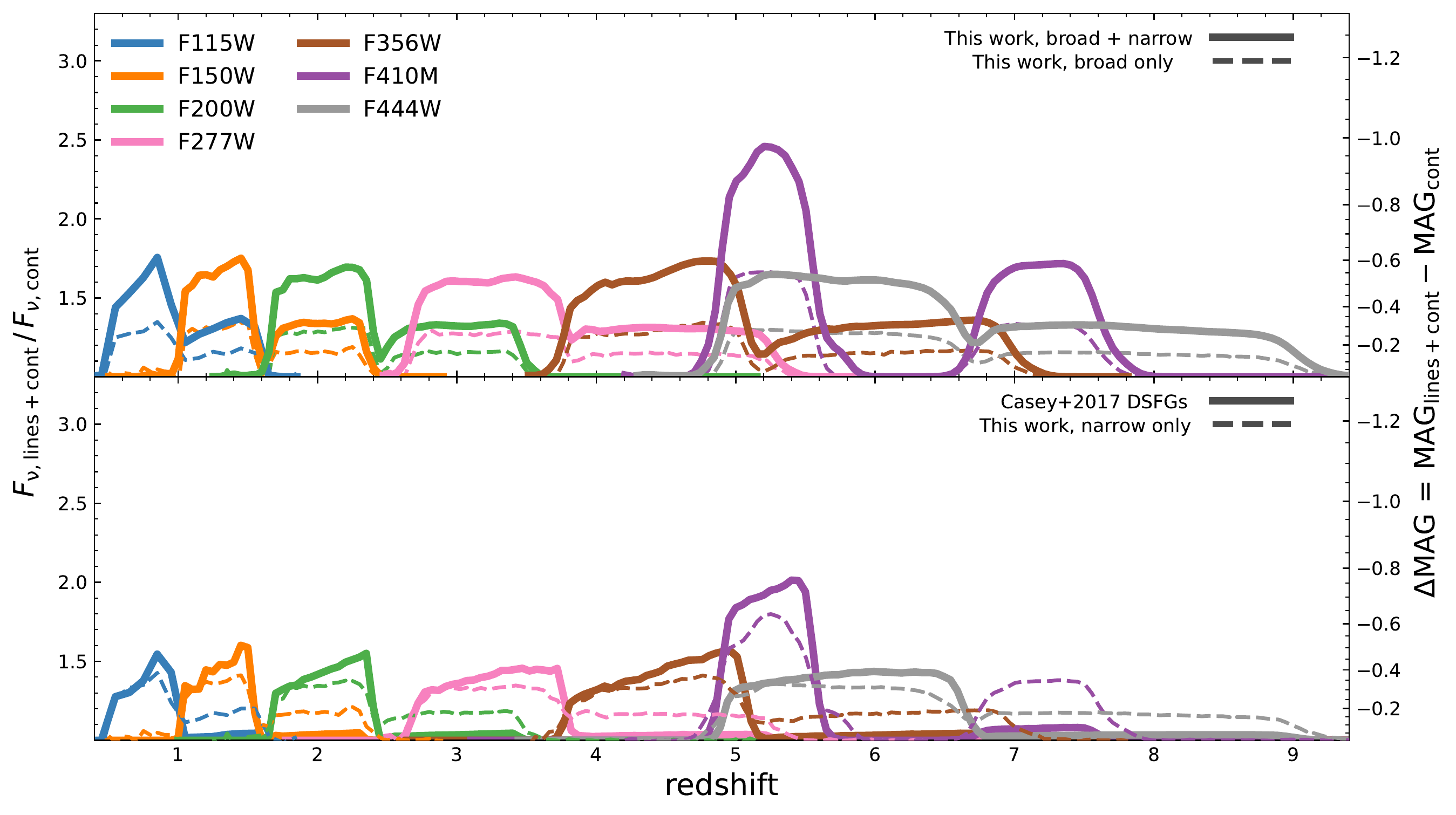}
    \caption{Increase in \textit{JWST}/NIRCam flux by strong rest-frame emission lines for the average SED of hot, dust-obscured and IR-lumionous galaxies between $\lambda_{rest}=0.34-0.8\,\mu$m as a function of redshift. (\textit{Top}) 
    %Shaded regions correspond to uncertainty on the flux ratio. 
    Solid lines account for broad and narrow velocity components, whereas dashed lines include only the broad component.  Maximally, strong nebular emission lines can boost the broadband flux between $\sim25-80\%$ ($|\Delta$MAG$|\,=0.2- 0.6$) from $z\sim1-8$. Medium-band filters such as F410M can be boosted by up to a factor of $2.5$ when they overlap with strong emission lines at $z\sim5.5$. The increase in flux attributed to velocity-broadened features is $\sim$25\% on-average ($|\Delta$MAG$|\,=0.2$).
    The double-peak effect for a given filter arises from the H$\alpha$ complex first passing through, followed by [OIII]+H$\beta$. 
    (\textit{Bottom}) Increase in flux attributed to strong line emission for the average DSFG spectrum of \citealt{Casey2017b} scaled to an H$\alpha$ star-formation rate of 100 $\mathrm{M_\odot\,yr^{-1}}$ (solid) and the narrow velocity component in our stack (dashed). The increase in flux by strong H$\alpha$+[NII] in the DSFG stack is consistent with the narrow line component for these lines the hot DOG stack. At $z\sim5$, all of the NIRCam LW filters are boosted by $\sim20\%-100\%$ for hot DOGs and DSFGs.}
    \label{fig:flux_ratio}
\end{figure*}

\subsection{Stacking}
Prior to stacking the data, we convert the observed wavelength range of each spectrum to the rest-frame with spectroscopic redshifts derived from optical lines with low errors ($\Delta z\sim10^{-3}$,  \citealt{Finnerty2020}). Next, we rebin the spectra to a common wavelength grid corresponding to the lowest rest-frame spectral resolution (R\,$\sim6400$). Finally, we calculate the sigma-clipped mean continuum flux from line-free regions which we use to normalize each spectra in the stack. 

We tested multiple stacking procedures and found that a mean noise-weighted continuum stack produced the cleanest continuum and highest line SNRs. In the stack, the input spectrum is first normalized by its sigma-clipped mean flux and then weighted by the spectral uncertainty per channel. This ensures that the stack is not dominated by particularly noisy spectral regions and/or the brightest spectra. As the goal of this analysis is the relative contribution of strong emission lines to photometry, we are not concerned with absolute normalization of the spectrum. To quantify the uncertainty on the continuum and line profiles, we repeat the stacking analysis 1000 times, using in each iteration 21 random samples of the input spectra with replacement (``bootstrapping''). From the bootstrapped uncertainties we determine that our final stacked spectrum is reliable between $\lambda_{\mathrm{rest}}=0.34-0.8\,\mu$m. 

The final stacked spectrum is shown in Figure \ref{fig:stack}. We detect strong [OIII], H$\beta$, [NII], H$\alpha$ emission lines, as well as [S II], [O II], [O I], [Ne III] and H$\gamma$. The [O I]$_{6300}$/H$\alpha$ line ratio is $0.5\pm0.1$, which is on the high end of the distribution measured for Seyfert galaxies in the \textit{Swift}-BAT AGN Spectroscopic Survey \citep{Koss2017}. 
EWs for the strong lines around the [OIII], H$\beta$ and [NII], H$\alpha$ complexes are listed in Table \ref{tab:ews}. Quoted uncertainties correspond to the standard deviation of EWs measured for 100 realizations of the spectrum perturbed by the spectral uncertainty per channel, and assuming a 10\% error on the continuum (uncertainties increase by a factor of $2.3$ assuming a 20\% error on the continuum). %These EWs are close to the average of EWs reported by \cite{Finnerty2020} for the hot DOGs individually.  

We also compare our stacked spectrum to the stack from \citealt{Casey2017b} derived from $20$ MOSFIRE spectra of DSFGs.\footnote{Available at \url{http://www.as.utexas.edu/~cmcasey/downloads.html}} %We find good qualitative agreement between the two stacks. 
The stack of \citealt{Casey2017b} exhibits narrower emission lines than our spectrum, and does not contain the broad outflow signatures founds in hot DOG rest-frame optical spectra \citep{Finnerty2020,Wu2018}. 
%One concern is that our stack averages out optical line ratio sensitivity to redshift driven by main-sequence, mass-metallicity and/or dust attenuation evolution. While the [O III] and H$\beta$ EWs of $z<3$ NIRES spectra are on-average lower than those at $z>3$ by a factor of $\sim1.5-2$ \citep{Finnerty2020}, both are consistent within $1\sigma$ because of substantial uncertainties for any one spectrum.

%\subsection{Optical line subtraction}

%\begin{figure}
%    \centering
%    \includegraphics[width=0.48\textwidth]{line_zoomin_nires.pdf}
%    \caption{Broad rest-frame optical emission lines in the stacked spectrum of $z\sim1-4$ IR-luminous galaxies detected with Keck II/NIRES.}
%    \label{fig:linesub}
%\end{figure}

\subsection{Synthetic Photometry}

To test the effect of emission lines on \textit{JWST}/NIRCam photometry, we subtract strong spectral features from the stack to produce a line-free continuum spectrum. To do so, we subtract the gaussian model fits from the stack. These lines include all shown in Figure \ref{fig:stack} (B, C) and include the range of velocity components required to fit individual hot DOGs, namely: broad [O III] and H$\alpha$ emission, narrow [O I], [O III], [S II], H$\alpha$, and H$\beta$ \citep{Finnerty2020}. 
We do not mask out [OII], [NeIII] and H$\gamma$ in this exercise as their lower EWs correspond to significantly less increase in broadband fluxes relative to [OIII]+H$\beta$ and [NII]+H$\alpha$. 
Following \citealt{Finnerty2020}, we assume narrow and broad profiles across different lines arise from the same kinematic components. This amounts to fixing  [N II] widths to that of the corresponding H$\alpha$ component. We also fix the [N II]$\lambda6548,\,\lambda6584\,$\AA\ ratio to 0.338 and the [O III]$\lambda4959,\,\lambda5007\,$\AA\ ratio to 0.335.
%The [O III]+H$\beta$ and [NII]+H$\alpha$ features which we mask are shown in Figure \ref{fig:stack} B and C, as well as the continuum level used to estimate the flux in the line-free stack. %We need broad and narrow velocity components to fit the average spectrum, as expected from the broad features found in individual spectra attributed to outflowing ionized gas \citep{Finnerty2020}.
In addition to the line-free stacked spectrum, we also compute a broad-line only spectrum (continuum $+$ broad line emission).

We calculate synthetic \textit{JWST}/NIRCam photometry using the filter response profiles provided by the \textit{JWST} User Documentation. We convolve each filter with both the stacked spectrum and line-subtracted stack for a range in redshift between $z=1-9$ in steps of $\Delta z=0.05$. We then take the ratio of filter flux between the line stack and line-subtracted (or broad line only) stack to infer the increase in flux attributed to emission lines as a function of redshift. 

\subsubsection{Composite DSFG spectrum from \cite{Casey2017b}\label{sec:c17}}
As both a check against our stack and a test for systems at lower \lir\ than the hot DOGs, we repeat our synthetic photometry calculations for the composite DSFG spectrum from \citealt{Casey2017b}. We scale their continuum-subtracted H$\alpha$ flux in their stack to the equivalent of 100 $\mathrm{M_\odot\,yr^{-1}}$ in star-formation rate 
using the F$_\mathrm{H\alpha}$ calibration of \cite{Murphy2011}. As the change in flux density due to nebular emission is a function of the relative strength between lines and continuum, we add the scaled DSFG spectrum to the empirically-derived rest-frame $0.1-1\,\mu$m mean DSFG SED from \citealt{Casey2014b}.
%add the continuum-subtracted and scaled DSFG stack to the average rest-frame $0.1-1\,\mu$m empirical SED template from \citealt{Casey2014b}.
For the line-free calculation we simply mask H$\alpha$, [O III] and H$\beta$ from the stack prior to performing synthetic photometry, equivalent to computing fluxes for the continuum DSFG SED without adding the lines. 

\begin{deluxetable}{lcr}
\tabletypesize{\small}
\tablewidth{0pt}
\tablecolumns{3}
\tablecaption{Strong optical emission line characteristics in our stacked spectrum \label{tab:ews}}
%\tablehead{\colhead{Line}&\colhead{EW}&\colhead{FWHM}&\colhead{}&\colhead{[\AA]}&\colhead{[km s$^{-1}$]}}
\tablehead{\colhead{Line} & \colhead{EW (\AA)} & \colhead{FWHM (km s$^{-1}$)}}
\startdata
$\mathrm{H}\beta$                       & $45\pm12$       &  $1450$  \\ 
$\mathrm{H}\alpha_{\mathrm{narrow}}$    & $222\pm27$      &  $730$   \\
$\mathrm{H}\alpha_{\mathrm{broad}}$     & $540\pm80$      & $4000$   \\ 
$\mathrm{[O III]_{5007}}$               & $127\pm19$      &  $1000$  \\ 
$\mathrm{[O III]_{4959}}$               & $43\pm8$        &  $1000$  \\ 
$\mathrm{[O III]_{5007,broad}}$         & $144\pm49$      & $7300$   \\ 
$\mathrm{[O III]_{4959,broad}}$         & $48\pm32$       & $7300$   \\ 
$\mathrm{[O I]_{6300}}$                 & $109\pm20$      &  $1600$  \\ 
$\mathrm{[O I]_{6363}}$                 & $38\pm13$       &  $1600$  \\ 
$\mathrm{[N II]_{6548}}$                & $35\pm8$        &  $730$   \\ 
$\mathrm{[N II]_{6583}}$                & $102\pm17$      &  $730$   \\ 
$\mathrm{[S II]_{6716}}$                & $103\pm18$      & $1200$   \\ 
$\mathrm{[S II]_{6730}}$                & $43\pm11$       & $1200$   \\ 
\hline
$A_V\,(\mathrm{H\alpha_{narrow}\,/H\beta})$ & $6\pm1$ & \\
$A_V\,(\mathrm{H\alpha_{tot}\,/H\beta})$ & $10\pm1$ & \\
\enddata
\end{deluxetable}

\begin{figure*}[ht!]
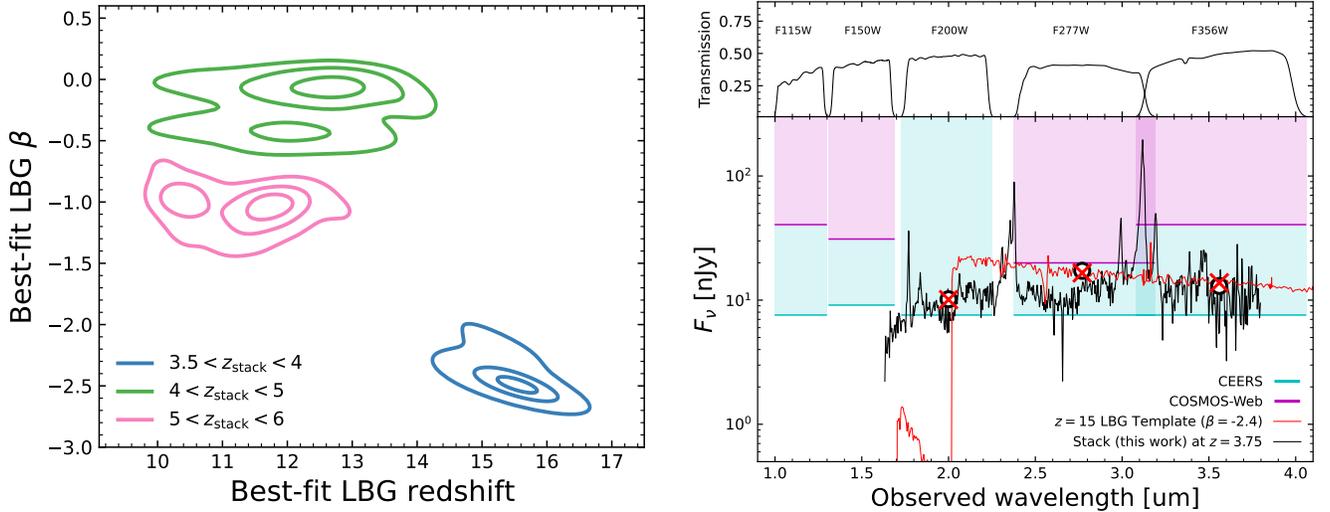

\gridline{\fig{z_beta_posts.pdf}{0.5\textwidth}{}
          \fig{sedfit_dsfg3_lbg15.pdf}{0.48\textwidth}{}
          }
          \vspace{-24pt}
\caption{
(\textit{Left}) Allowed LBG redshift and UV slope $\beta$ solutions when fitting the synthetic NIRCam flux densities of our stack redshifted to $z=3.5-4$ (blue), $z=4-5$ (green), and $z=5-6$ (pink). Posterior contours are drawn at the 16th, 50th, and 84th percentiles.  At $z>4$ [NII]+H$\alpha$ increase the NIRCam flux density more so than [OIII]+H$\beta$ which does not allow the strong lines to mask the red continuum and therefore precludes LBG solutions with $\beta<-1.5$. Between $3.5<z<4$ [NII]+H$\alpha$ falls within F277W while [OIII]+H$\beta$ is missed by F200W, allowing degenerate solutions with blue $\beta\leq-2$ LBGs at $z\sim14-17$. (\textit{Right}) Illustration of the degeneracy between $z\sim15$ candidates and $z<4$ dusty galaxies with strong rest-frame optical/emission lines. In this example, we redshift our stacked spectrum to $z=3.75$ where strong line emission boosts the F277W filter flux by 60\%. We then compute F200W, F277W, and F356W \textit{JWST}/NIRCam photometry (black circles), assuming non-detections in F115W and F150W. % (black arrows) as would be expected for $z>10$ candidates, 
We fit the synthetic photometry from the stack (circles) with an LBG template (red), deriving a photometric redshift of $z_{phot}=15$ and UV spectral index $\beta=-2.4$. Strong emission lines mask the red slope of the dusty template between F277W and F356W, and the SED is further confused with the Lyman break falling halfway between F200W. Such scenarios are possible given the relative filter depths of \textit{JWST} Cycle 1 NIRCam extragalactic surveys in CEERS (blue, \citealt{Bagley2022,Finkelstein2022b}) and COSMOS-Web (pink, \citealt{Casey2022}) for example.
}
\label{fig:confusion}
\end{figure*}

\section{Results and Discussion\label{sec:discussion}}

The results of our synthetic photometry are shown in Figure \ref{fig:flux_ratio}, which gives the flux ratio between our fiducial and line-subtracted stacked spectrum % of $z=1-4$ dusty galaxies 
and the DSFG stack from \cite{Casey2017b}. 
For the former, we also show the increase in flux seperated between the narrow and broad velocity components. 
On average, strong narrow$+$broad rest-frame optical lines increase NIRCam fluxes by factors of $\sim1.2-1.7$, with corresponding change in apparent magnitude by $0.2-0.6$. The maximal increase in flux %persists over a redshift range of $\Delta z\sim0.4-1$ 
occurs when any particular wide-band filter is centered on the strong [NII]+H$\alpha$ complex. The [OIII] and H$\beta$ lines collectively increase the wide-band flux maximally by $\sim$20\%. Their broad components and those of [NII] and H$\alpha$ increase synthetic flux densities by $1.2\times$ on average, accounting for $\sim66\%$ of the boost for [OIII]+H$\beta$ and $25\%$ for [NII]+H$\alpha$. Medium-band filters are more affected by the presence of strong emission lines %over $\Delta z\sim1$ 
and can be dominated by factors of $\sim2.5$ ($1$ mag) by emission lines when redshifted to the line's rest wavelength. 
For example, the F410M flux is increased by a factor of $2.5$ at $z=5.5$. In fact, $z=5$ is a special regime where a boost in flux density is seen for all NIRCam LW filters. While we do not show the increase in flux attributed to the relatively weaker [OII] line on Fig.~\ref{fig:flux_ratio}, this effect is maximally $\sim10\%$ if we mask the line following the methods outlined for [OIII]+H$\beta$ and [NII]+H$\alpha$.

%Comparable to the results for the hot DOG stack,
The strong lines in the DSFG stack from \citealt{Casey2017b} which we have scaled to an H$\alpha$ SFR of $100\,\mathrm{M_\odot\,yr^{-1}}$ (see section \ref{sec:c17})  increase broadband fluxes by up to a factor of $\sim1.5$. Such boosting occurs over similar ranges in redshift and to the same degree as found for the narrow line components in the hot DOG stack. This demonstrates that significant line contamination can be present in the NIRCam photometry for IR-luminous galaxies more normal than the relatively extreme hot DOGs.

Given extreme levels of attenuation in massive dust-obscured galaxies at high-redshift, their rest-frame optical spectra contain a combination of significantly reddened continuum with $\lesssim5\%$ of the total un-obscured light escaping from the least obscured regions \citep{Chapman2005,Howell2010}. With a combination of strong lines emergent from less obscured regions on top of the very red continuum, $\sim0.1-1\,\mu$m photometry of dusty galaxies can mimic that of ultra-high redshift LBG candidates in large surveys \citep{Zavala2022,Fujimoto2022}. To quantify the parameter space where this confusion is significant, we fit an LBG template to the synthetic NIRCam flux densities derived from our stacked spectrum. We first normalize the stack to a continuum flux on the order of $\sim10$ nJy over $\lambda_{obs}=2-3.5\,\mu$m, and assume it to be undetected in F115W and F150W. This represents a plausible scenario given the relative filter depths of \textit{JWST} Cycle 1 extragalactic deep fields \citep{Bagley2022,Finkelstein2022b,Casey2022}, %, and one which would imply a dropout redshift of $z>10$. 
and is similar to CEERS-93316 \citep{Donnan2022} (CEERS2\_2159, \citealt{Finkelstein2022b}) $-$ a $z=16.4$ LBG candidate selected from CEERS \citep{Finkelstein2022,Bagley2022}. CEERS-93316 has a tentative 2.6$\sigma$ SCUBA-2 detection \citep{Zavala2022} and environmental evidence \citep{Naidu2022a} both indicating a possible lower redshift solution at $z\sim4.8$.

Figure \ref{fig:confusion} (\textit{Left}) shows the 2D posterior distribution in redshift and UV slope $\beta$ for LBG template fits to our stacked spectrum's synthetic NIRCam flux densities. We repeat the fitting analysis 1000 times after perturbing the input spectrum by the spectral uncertainty, and in three redshift ranges for the stack: $3.5<z<4$, $4<z<5$, and $5<z<6$. The cumulative EW of H$\alpha$+[NII] is greater than EW([OIII]+H$\beta$) by a factor of $\sim3$ which precludes LBG fits with $\beta<-1.5$ when the stack is redshifted to $z>4$. This is because both features fall within a broadband filter and so the strong lines do not mask the red continuum in the stack. At $3.5<z<4$ F277W picks up the strong H$\alpha$+[NII] emission while [OIII]+H$\beta$ is missed by F200W. This produces degenerate photometry with $z\sim16$ LBGs. At $z\sim16$, the Lyman break falls halfway between F200W mimicking the red slope of the dusty galaxy stack while the very blue continuum mimics the F277W flux density of the line-contaminated stack. 
In summary, the hot DOG stack can reproduce very blue UV slopes $\beta\sim-2.5$ for $z_{\mathrm{stack}}\sim3.5-4$ but not for $z_{\mathrm{stack}}>4$. This supports the purity of the very blue NIRCam samples of \citealt{Cullen2022} and \citealt{Topping2022}, which predominantly have $\beta<-1.5$ and $7<z<14$.

As further demonstration of the confusion between our stack at $z\sim3.5-4$ and $z\sim16$ LBGs, we show in Figure \ref{fig:confusion} (\textit{Right}) the LBG fit to our stack's \textit{JWST}/NIRCam flux densities. At $z_{\mathrm{stack}}=3.75$ we find a best-fit LBG solution with $z=16$ and UV spectral index $\beta=-2.4$. Lower redshift ($z<4$) solutions with red continuum slopes and flux densities dominated by strong emission lines should be considered when fitting the very blue ($\beta<-2$) spectral energy distributions (SEDs) of $z\sim16$ candidates. The lower redshift solutions could be ruled out with medium-band filters, longer wavelength sampling using NIRCam's redder filters, MIRI observations, and/or far-IR/sub-mm follow-up to detect cold dust continuum and fine-structure lines \citep{Fujimoto2022}. Measurements that strongly rule out UV spectral indices $\beta<-1.5$ and only allow lower$-z$ solutions at $z>4$ should be particularly constraining against massive, IR-luminous interlopers with strong optical lines provided they sample the SED with more than three filters. %and over the range of Balmer lines at $z\sim3-6$

Based on the first analysis of \textit{JWST} deep field observations at 5$\sigma$ point-source depths between $\sim28-29$ MAG, the projected sky density of candidates at $z>10$ is approximately $350\,\mathrm{\pm120}\,\mathrm{deg^{-2}}$ %$7.5\times10^{-5}\,\mathrm{Mpc^{-3}}$ at $M_\mathrm{UV}>-22$
\citep{Donnan2022,Finkelstein2022,Naidu2022,Harikane2022}. Although preliminary, these source counts represent the population which could potentially be contaminated by low$-z$ dusty interlopers. In contrast, the sky density of luminous IR galaxies with \loglir$\,>12\,(12.5)$ and $z\sim3-4$ is $400\,\mathrm{deg^{-2}}$ ($100\,\mathrm{deg^{-2}}$) \citep{Casey2018,Zavala2021}. If we assume the samples of \cite{Finnerty2020} and \cite{Casey2017b} include a range of physically possible rest-frame optical properties for IR-bright galaxies (\loglir$\,>11$), then their similar number counts to ultra high-redshift LBG candidates may be reason to be concerned about contamination.
%then to first order this suggests a non negligible contribution of $z<6$ dusty interlopers to the ultra high-redshift LBG catalogs by virtue of their similar number counts to the known populations of $4<z<6$ dusty galaxies. %Moreover, we only account for the IR-bright (\loglir$\,>12$) population in this calculation, while 
The fainter dusty galaxy population with \loglir$\,<11$ are much more numerous based on the general shape of 1 mm number counts \citep{Fujimoto2016,GonzalezLopez2020}, and may also be an important source of contamination as galaxies fainter in the IR are less likely to be significantly obscured in the rest-frame optical. 
%and strong narrow emission lines can increase NIRCam fluxes by $\sim50-100\%$ in the LIRG and ULIRG range (Fig.~\ref{fig:flux_ratio}, \textit{bottom}). 
Further spectroscopic follow up is required to assess the purity of ultra-high redshift catalogs. In the meantime, F150W dropouts ($z>10$) with $\beta\sim-2$ and no/poor SED constraint above NIRCam/F356W should be checked against possible intermediate-redshift dusty galaxy solutions. 

%are more likely to be intermediate-redshift DSFGs than bluer F150W dropouts.

%\begin{figure}
%    \centering
%    \includegraphics[width=0.48\textwidth]{sedfit_dsfg3_lbg15.pdf}
%    \caption{Illustration of the degeneracy between $z\sim15$ candidates and $z<4$ dusty galaxies with strong rest-frame optical/emission lines. In this example, we redshift our stacked spectrum to $z=3.75$ where strong line emission boosts the F277W filter flux by 60\%. We then compute F200W, F277W, and F356W \textit{JWST}/NIRCam photometry (black circles), assuming non-detections in F115W and F150W. We fit the synthetic photometry from the stack (circles) with an LBG template (red), deriving a photometric redshift of $z_{phot}=15$ and UV continuum spectral index $\beta=-2.4$. Strong emission lines mask the red slope of the dusty template between F277W and F356W, and the SED is further confused with the Lyman break falling halfway between F200W. Such scenarios are possible given the relative filter depths of \textit{JWST} Cycle 1 NIRCam extragalactic surveys in CEERS (blue, \citealt{Bagley2022,Finkelstein2022b}) and COSMOS-Web (pink, \citealt{Casey2022}) for example.    }
%    \label{fig:lbgfit}
%\end{figure}

\section{Summary and Conclusion\label{sec:conclusion}}

In this Letter we test the response of \textit{JWST} NIRCam filters over broad rest-frame optical emission lines in the average spectrum of hot, dust-obscured galaxies at $z\sim1-4$ and dusty, star-forming galaxies. As an empirical approach, we stack a sample of $21$ IR luminous galaxies with rest-frame optical spectra from Keck II/NIRES which we then compute synthetic photometry for between $z=1-9$. Our main results are as follows: 
\begin{enumerate}
    \item We measure broad rest-frame optical emission lines in the stack of $z\sim1-4$ hot, dust-obscured galaxies. In particular, we measure [OIII] and  H$\alpha$ EWs between $100-500$ \AA\ which are high relative to normal star-forming galaxies at high-redshift.%$z<3$ AGN and Type II QSOs. 
    \item After masking out strong emission features from the spectrum, we measure synthetic NIRCam photometry with and without the lines. Narrow and broad components for [OIII] and H$\beta$ increase the measured filter flux by $30\%$, and H$\alpha+$[NII] by $60\%$ on average. %over redshift windows with $\Delta z\sim0.4-1$. 
    Narrowband filters such as F410M can have their flux increased by a factor of $2-3$ ($0.7-1.2$ MAG). 
    \item Rest-frame optical photometry of dusty galaxies with strong nebular lines at $z\sim3.5-4$ could be important contaminants in F150W dropout ($z>10$ candidate) catalogs as the strong lines can help mask red UV spectral indices. However, UV spectral indices $\beta<-1.5$ are difficult for our stacked spectrum to reproduce for interloper redshifts $z>4$.%, and in some cases the H$\alpha$+[NII] feature can be identified by enhanced narrowband filter flux. 
\end{enumerate}
Distinguishing between different galaxy populations with \textit{JWST} imaging is a key first step towards testing various aspects of galaxy formation. %from the onset of the first galaxies to the assembly of massive ones. 
While this work has focused just on \textit{JWST}'s NIRCam filters, the inclusion of deep MIRI photometry extending to longer wavelengths will add significant constraint on various redshift solutions to photometric fitting codes. In the absence of high SNR coverage in redder filters, far-IR/sub-mm followup can help identify dusty galaxies. On the near horizon, ToLTEC on the Large Millimeter Telescope (LMT) Alfonso Serrano will map multiple extragalactic fields (COSMOS, UDS, GOODS-S) down to the LIRG limit at 1.1, 1.4, and 2 mm as part of ``The TolTEC Ultra-Deep Galaxy Survey,'' a public legacy program. These public data sets are well suited to quickly identify sub-mm bright DSFG counterparts to \textit{JWST} sources. 

\begin{acknowledgments}
L.F. is a member of Student Researchers United (SRU-UAW). The data presented herein were obtained at the W. M. Keck Observatory, which is operated as a scientific partnership among the California Institute of Technology, the University of California, and the National Aeronautics and Space Administration. The Observatory was made possible by the generous financial support of the W. M. Keck Foundation. We wish to acknowledge the critical importance of the current and recent Maunakea Observatories daycrew, technicians, telescope operators, computer support, and office staff employees, especially during the challenging times presented by the COVID-19 pandemic. Their expertise, ingenuity, and dedication is indispensable to the continued successful operation of these observatories. The authors wish to recognize and acknowledge the very significant cultural role and reverence that the summit of Maunakea has always had within the indigenous Hawaiian community. We are most fortunate to have the opportunity to conduct observations from this mountain.
\end{acknowledgments}

\clearpage
\bibliography{references.bib}

\end{document}